\documentclass{elsart}

\usepackage{epsfig}

\newcommand{\bea}{\begin{eqnarray}}
\newcommand{\eea}{\end{eqnarray}}
\newcommand{\be}{\begin{equation}}
\newcommand{\ee}{\end{equation}}

\begin{document}


\begin{flushright}
SNUTP 99-017\\
hep-ph/yymmdd\\
March 1999
\end{flushright}


\begin{frontmatter}

\title{ 
Probing anomalous right--handed current in $Zb\bar{b}$ vertex
}

\author[ctp]{Kang Young Lee\thanksref{elee}}
\thanks[elee]{\tt kylee@ctp.snu.ac.kr}
\address[ctp]{Center for Theoretical Physics, Seoul National
University, Seoul 151--742, Korea}

\begin{abstract}

Motivated by $-2 \sigma$ deviation 
from the Standard Model prediction
for the forward--backward asymmetry of $Z \to b \bar{b}$ decay 
in the LEP data, we explore the anomalous couplings of $Z b \bar{b}$
vertex through the electroweak precision test
extended by including $A_{FB}^b$.
We introduce a new variable $\epsilon'_b$ 
to measure the anomalous right-handed coupling 
in $Z b \bar{b}$ vertex.
Implications of our analysis on the nonuniversal contact
interactions are studied and
the left-right model is also considered as a possible
origin of the anomalous right-handed couplings.
\\
\\

\end{abstract}

\begin{keyword}
forward--backward asymmetry, 
$Z b \bar{b}$ vertex,
the anomalous couplings,
the nonuniversal interactions,
the left-right model
\end{keyword}

\end{frontmatter}



\newpage

\section{Introduction}

The Standard Model (SM) has accomplished a great success
to describe enormous phenomena of elementary particles.
Most of the experimental data have shown quite a good
quantitative agreement with the SM predictions 
for various observables. 
Precision measurements of the $Z$--pole observables 
at LEP and SLD have provided highly accurate tests 
on the Standard Model \cite{sm1,sm2,sm3}.
After the controversies about the discrepancy of $R_b$
($\equiv \Gamma(Z \to b \bar{b}) /\Gamma(Z \to \mbox{hadrons})$)
are resolved along with the improvement of the experiment, 
no evidences of the new physics signal from colliders are reported yet.
The new physics effects are thought
to be as large as the loop effects of the SM at most.

We are still looking for a hint of the discrepancy 
from the SM predictions in the list of the LEP and SLC data.
The forward--backward asymmetry of $Z \to b \bar{b}$ decay, $A_{FB}^b$,
may be a clue of such discrepancies
as it shows $-2 \sigma$ deviation from the SM prediction.
If we consider the left-right forward-backward asymmetry $A_b$,
which is directly measured by SLD and is related to $A_{FB}^b$,
the discrepancy is larger.
The combined fit for the LEP and SLD measurements gives
$A_b = 0.881 \pm 0.018$ which is $3 \sigma$
away from the SM prediction \cite{sm3}.
Taking this deviation to be serious,
anomalous couplings are required in the $Z b \bar{b} $ vertex.
At the same time we demand that the partial decay width $\Gamma_b$ 
(alternatively $R_b$) should be kept within 
the experimental bound with these anomalous couplings.

The possibility of anomalous couplings are discussed 
in many literatures. 
Field presented an detailed analyses on the asymmetries
and couplings in  model independent way \cite{field}.
Chang and Ma suggested a vectorlike heavy quark model
to explain $A_{FB}^b$ and $R_b$ data \cite{chang}.
In this letter, we attempt to extract the anomalous
couplings in $Z b \bar{b}$ vertex
which explain $A_{FB}^b$ data of LEP and SLD together with $R_b$
through an extended scheme of electroweak precision test.
The remarkable feature is that the anomalous right-handed coupling
is required as well as the left-handed one 
since variables for asymmetries are affected by the ratio
of the right-handed coupling to the left-handed one.
Since the SM electroweak radiative correction is almost
left-handed, the anomalous right-handed current may be a sensitive
probe to new physics effects.
Altarelli et al. have suggested the $\epsilon$ variables
for the electroweak precision test \cite{altarelli1}
and modified the analysis including $\Gamma_b$ 
to describe the large $m_t^2$ dependences 
of electroweak radiative corrections to $Z b \bar{b}$ vertex
\cite{altarelli2}.
However it is still dissatisfactory to describe the general couplings. 
Here we suggest an extension including $A_{FB}^b$ data 
to extract the general type of the anomalous couplings of 
$Z b \bar{b}$ vertex.

We apply our analysis to the model 
in the presence of the nonuniversal contact interactions.
The nonuniversal interaction acting on the third generation can be 
an attractive candidate for the new physics
\cite{hill1,zhang,lee,hill2,hill3},
since we favor that the SM predictions for other flavours
should not be much disrupted by the new physics effects.

As another realization of the new physics effects, 
we consider the left-right model (LR model) in the general scheme.
In the framework of LR model
based on the extended electroweak gauge group
SU(2)$_L \times$SU(2)$_R \times$U(1),
the anomalous right-handed couplings are naturally introduced 
as well as the left-handed ones as is desirable.
In this model the anomalous couplings in $b$ sector are
directly related to those in the lepton sector
and should be strictly constrained.
For given bounds of the LR model parameter set,
both of the $A_{FB}^b$ and the $R_b$ data 
can be shown to be accommodated.

This paper is organized as follows:
We extract the anomalous current interactions 
in $Z b \bar{b}$ vertex in terms of the model independent 
parameterization of the basic observables including $A_{FB}^b$ 
in Section II. 
Our analysis in terms of new $\epsilon$ variables are applied
to the minimal nonuniversal contact term that is $d>4$ 
and its effect on the $Z b \bar{b}$ vertex
in Section III.
The model with SU(2)$_L \times$SU(2)$_R \times$U(1)
gauge group is considered in Section IV.
We present constrained parameter set $(\epsilon_b,\epsilon'_b)$ 
and show that both of the $A_{FB}^b$ and the $R_b$ data 
can be accommodated.
Finally we conclude in Section V.

\section{Model independent parameterization}

Provided that we allow the additional contribution 
of the new physics to the $Z b \bar{b}$ vertex,
we write the most general amplitude for $Z \to b \bar{b}$ decay
for the model--independent analysis as 
\be
M(Z \to b \bar{b}) = \frac{g}{2 \cos \theta_W}
     \epsilon^\mu \bar{u} \big( \gamma_\mu (g_{bV}^0 - g_{bA}^0 \gamma_5)
    + \Delta^L_b \gamma_\mu P_L + \Delta^R_b \gamma_\mu P_R \big) u~,
\ee
where $\epsilon_\mu$ is the polarization vector for $Z$ boson and
$g_{bV}^0$ ($g_{bA}^0$) are tree level vector 
(axial--vector) couplings of $b$ quark pair to $Z$ boson given by:
\be
g_{bV}^0 = -\frac{1}{2} + \frac{2}{3} s_0^2~,~~~~~
g_{bA}^0 = -\frac{1}{2}~,
\ee
with the Weinberg angle at tree level, $s_0^2$, satisfying 
$s_0^2 c_0^2 = \pi \alpha(m_{_Z}) /\sqrt{2} G_F m_{_Z}^2$.
In the SM, $ \Delta^L_b = \Delta^{SM}_b (m_t^2)$ and $ \Delta^R_b=0$
where the leading contribution of $\Delta^{SM}_b(m_t^2)$
in the large $m_t$ limit is given by \cite{akhundov}
\be
\Delta^{SM}_b(m_t^2) = \frac{\alpha}{4 \pi \sin^2 \theta_W}
          | V_{tb} |^2 \frac{1}{2} \left(
          \frac{m_t^2}{m_{_W}^2} + 
           \left( \frac{8}{3} + \frac{1}{6 \cos^2 \theta_W} \right) 
                     \log \frac{m_t^2}{m_{_W}^2} \right)~,
\ee
which arises from the top quark exchange diagrams
shown in Fig. 1.

\begin{figure}[t]
\begin{center}
\epsfig{file=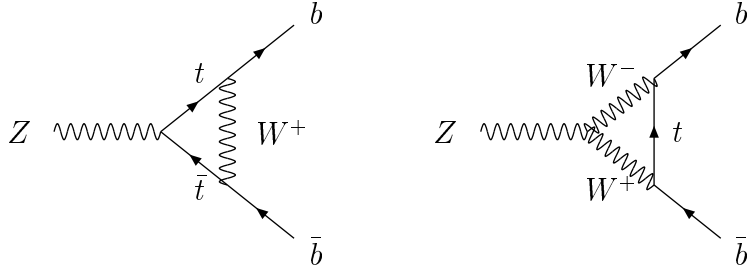,width=10cm}
\caption{ Electroweak radiative corrections to $Z b \bar{b}$ vertex.
}
\end{center}
\vskip 0.5cm
\end{figure}

We present the analysis in terms of the precision variables 
in order to incorporate the SM corrections.
It is helpful to introduce the precision variables
for the study of the new physics effects on the electroweak data
because the additional contributions of new physics are expected 
to be comparable with the loop contributions of the SM.
The $\epsilon$ analysis suggested by Altarelli et al. 
\cite{altarelli1,altarelli2}
provides a model independent way to analyze the electroweak 
precision data.
In this scheme, the electroweak radiative corrections
containing whole $m_t$ and $m_H$ dependencies are parametrized
into the parameters $\epsilon$'s and thus the $\epsilon$'s can be
extracted from the data without specifying $m_t$ and $m_H$.

The universal correction terms to the electroweak form factors
$\Delta \rho$ and $\Delta k$ are defined from the vector and
axial vector couplings of lepton pairs to $Z$ boson
as Eq. (4) of Ref. \cite{altarelli2}
and extracted from the inclusive partial decay width $\Gamma_l$
and the forward-backward asymmetry $A_{FB}^l$.
Thus we present the amplitude in terms of 
the vector and axial-vector couplings 
instead of the left- and right-handed couplings in Eq. (1).
Meanwhile the SM correction (3) is left-handed
and so we introduce the additional terms in the left-
and right-handed basis.
Another correction term $\Delta r_W$ is obtained
from the mass ratio $m_{_W}/m_{_Z}$ by the Eq. (1)
of Ref. \cite{altarelli2}.
The $\epsilon$ parameters are defined by the linear combinations
of correction terms to avoid the new physics effects being masked 
by the large $m_t^2$ corrections in $\epsilon_2$ and $\epsilon_3$.
We note that $\Delta r_W$ is irrelevant for our analysis and
affects only on $\epsilon_2$.
Hence we lay aside $\epsilon_2$ in our analysis of this letter.

The parameter $\epsilon_b$ is introduced to measure 
the additional contribution to the $Z b \bar{b}$ vertex 
due to the large $m_t$-dependent corrections in the SM.
Since the electroweak radiative correction of the SM
shown in Eq. (3) is left-handed in the large $m_t$ limit,
Altarelli et al. have defined $\epsilon_b$ 
through the effective couplings $g_{bA}$ and $g_{bV}$
in the following manner:
\bea
g_{bA} &=& -\frac{1}{2} (1+\frac{1}{2} \Delta \rho) (1+\epsilon_b)~,
\nonumber \\
x_b &\equiv& \frac{g_{bV}}{g_{bA}} = 
\frac{ 1-\frac{4}{3} \sin^2 \theta^l_{eff} + \epsilon_b}
     {1+\epsilon_b}~.
\eea
of which asymptotic contribution is given by 
$\epsilon_b \approx -G_F m_t^2/4 \pi^2 \sqrt{2}$.
In this expression, $ \sin^2 \theta^l_{eff}$ denotes
the effective Weinberg angle including the SM loop corrections
to the lepton sector.

However $\epsilon_b$ in Eq. (4) cannot be 
the most general deviations from the Standard Model 
of $Z b \bar{b}$ vertex and is not suitable
for incorporating new physics effects.
We introduce another parameter to describe the additional
right--handed current interaction effects.
Firstly we define $\Delta \rho_b$ by a deviation 
from the axial coupling of lepton sector:
\be
g_{bA} = g_{A} (1+\Delta \rho_b)
       = -\frac{1}{2} \left( 1+\frac{1}{2} \Delta \rho \right) 
          (1+\Delta \rho_b) ~.
\ee
Next we introduce another correction term $\Delta k_b$
analogous to the lepton sector as
\be
\sin^2 \theta^b_{eff} = \sin^2 \theta^l_{eff} (1+\Delta k_b)
= s_0^2 (1+\Delta k) (1+\Delta k_b)~,
\ee
which leads to
\be
x_b \equiv \frac{g_{bV}}{g_{bA}}
= 1-\frac{4}{3} \sin^2 \theta^b_{eff} 
= \frac{1-\frac{4}{3} \sin^2 \theta^l_{eff} - \Delta k_b}
       {1 - \Delta k_b}~.
\ee
The correction terms $\Delta \rho_b$ and 
$\Delta k_b$ are defined in accord with those of lepton sector.
To avoid being masked by the electroweak radiative corrections 
of the SM, we define the new epsilon parameters by the relations
\be
\epsilon_b \equiv \Delta \rho_b,~~~~~~~~
\epsilon'_b \equiv \frac{2}{3} s_0^2 (\Delta \rho_b + \Delta k_b),
\ee
with canceling $\Delta^{SM}_b(m_t^2)$ in $\epsilon'_b$.
In the SM limit, $ \Delta \rho_b =-\Delta k_b = \Delta_b^{SM}(m_t^2) $
and consequently $\epsilon'_b = 0$. 
Hence $\epsilon_b$ goes to the original definition
in the SM limit while $\epsilon'_b$ purely measures the
anomalous right-handed current interaction and goes to 0
in the SM limit.
The parameters $\epsilon_b$ and $\epsilon'_b$ are extracted
from the observables of the inclusive decay width $\Gamma_b$
and forward-backward asymmetry of $b \bar{b}$ production $A_{FB}^b$.
In consequence, the four parameters 
$\epsilon_1$, $\epsilon_3$, $\epsilon_b$, and $\epsilon'_b$
are set to be one to one correspondent to the observables
$\Gamma_l$, $A_{FB}^l$, $\Gamma_b$, and $A_{FB}^b$.
The quadratic $m_t$ dependences of the SM electroweak radiative
corrections appear in $\epsilon_1$ and $\epsilon_b$
while the $m_t$ dependence of $\epsilon_3$ is logarithmic.
The parameter $\epsilon'_b$ is identical to the anomalous
right-handed current interactions in $Z b \bar{b}$ vertex
and expected to be sensitive to the new physics.

\begin{figure}[t]
\centering
\epsfig{file=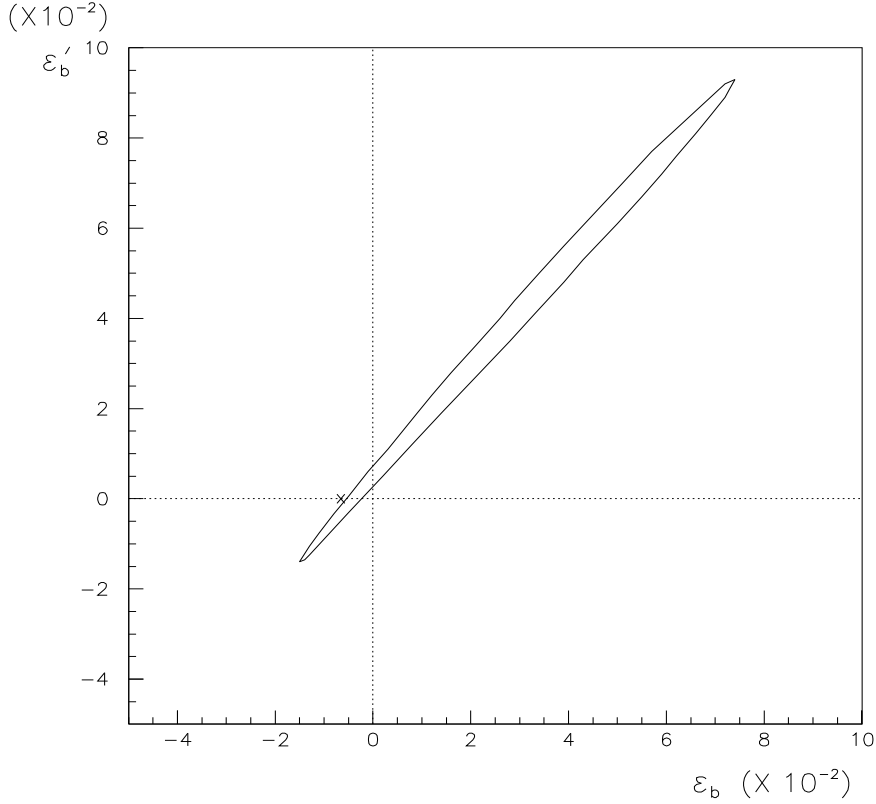,height=12cm,width=13cm}
\caption{ Available region in $\epsilon_b$ - $\epsilon'_b$ plain
obtained from experiment.
}
\end{figure}

We have the modification of the linearized relations
between the observables $\Gamma_b$, $A_{FB}^b$
and the epsilon parameters as
\bea
\Gamma_b &=& \Gamma_b |_B (1 + 1.42 \epsilon_1 - 0.54 \epsilon_3
+2.29 \epsilon_b - 1.89 \epsilon'_b)
\nonumber \\
A_{FB}^b &=& A_{FB}^b |_B (1 + 17.5 \epsilon_1 - 22.75 \epsilon_3
+0.157 \epsilon_b - 1.02 \epsilon'_b)
\eea
while the relations of $\Gamma_l$ and $A_{FB}^l$
remains intact as Eq. (123) of Ref. \cite{altarelli2}.
$\Gamma_b |_B$ and $ A_{FB}^b |_B $ are
the Born approximation values which are defined by the tree level
results including pure QED and pure QCD corrections and
consequently depend upon the values of $\alpha_s(m_{_Z}^2)$
and $\alpha(m_{_Z}^2)$.
The QCD corrections to the forward-backward asymmetries
of $b$ quark pair in $Z$ decays are given in the Ref. \cite{abbaneo}.
We obtain
$ \Gamma_b |_B = 379.8$ MeV, and $A_{FB}^b |_B = 0.1041$
with the values $\alpha_s(m_{_Z}^2) = 0.119$
and $\alpha(m_{_Z}^2) =1/128.90$.
We show the allowed region of $\epsilon_b$ and
$\epsilon'_b$ at 95 $\%$ C.L.
from the recent LEP$+$SLD data in Fig. 2.
The values of $m_t = 175$ GeV and $m_H=100$ GeV are used.
We obtain
\bea
\epsilon_b &=& (1.8 - 5.4) \times 10^{-2},
\nonumber \\
\epsilon'_b &=& (2.7 - 7.0) \times 10^{-2}.
\eea

\section{Nonuniversal contact interactions}

Models with nonuniversal contact interactions are
mainly motivated by the idea that mass of the top quark
is of order of the weak scale and so  the top quark
could be responsible for the electroweak symmetry breaking.
In general we have several contact terms which are $d>4$ at 
a high energy scale in those models.
We find a general list of contact terms in Refs. \cite{hill2,hill3}.
As a minimal contents of the model,
left-handed SU(2) doublet for the third generation and the right handed
singlet $t_R$ are coupled in a new gauge interaction.
The relevant term of the effective lagrangian is written by
\begin{equation}
L_{eff} = - \frac{1}{\Lambda} \bar{b} \gamma_{\mu} b
\bar{t} \gamma_{\mu} (f_V - f_A \gamma_5) t + ...~,
\end{equation}
where $f_V$ and $f_A$ are model parameters and 
$\Lambda$ is the new physics scale.
Following the Ref. \cite{hill1,hill3} for normalizations,
we define $f_A \sim 4\pi(0.11)$.
The effective corrections to $Z \to b \bar{b}$ decay as represented
in Eq. (1) are generated via the top quark loops given by
\begin{equation}
\Delta_b^{L,R} = \frac{N_c}{4 \pi^2} f_A \frac{m_t^2}{\Lambda^2} 
                 \ln \frac{\Lambda^2}{m_t^2}~,
\end{equation}
where $N_c = 3$.
Since $ \Delta_b^{L} = \Delta_b^{R}$, no additional contributions to 
$\epsilon_b $ and $\epsilon'_b = \Delta_b^R$. 
From the Fig. 2, we find that
$\epsilon'_b = (0 \sim -5) \times 10^{-3}$ 
when $\epsilon_b = \epsilon_b^{SM}$,
which yields the new physics scale
$\Lambda > 1.7$ TeV.

\begin{figure}[t]
\begin{center}
\epsfig{file=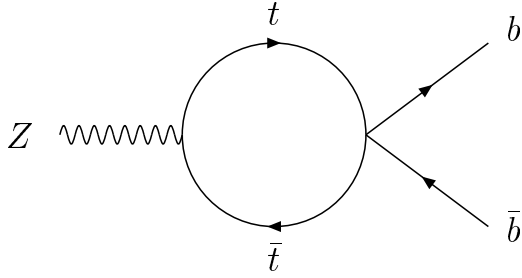,width=7cm}
\caption{ Vertex corrections to $Z \to b \bar{b}$ decay  
with the contact term.
}
\end{center}
\vskip 1cm
\end{figure}

\section{SU(2)$_L \times$SU(2)$_R \times$U(1) model}

If indeed the anomalous right-handed current exists 
in $Z b \bar{b}$ vertex, it demands to find an origin 
of such an anomaly as a next task.
Here we consider the left-right model (LR model) based on the
extended gauge group SU(2)$_L \times$SU(2)$_R \times$ U(1)
as a possible candidate.
We assume a general model without imposing the parity
on th lagrangian where the value of $g_R$ need not be equal to $g_L$.

\begin{figure}[t]
\epsfig{file=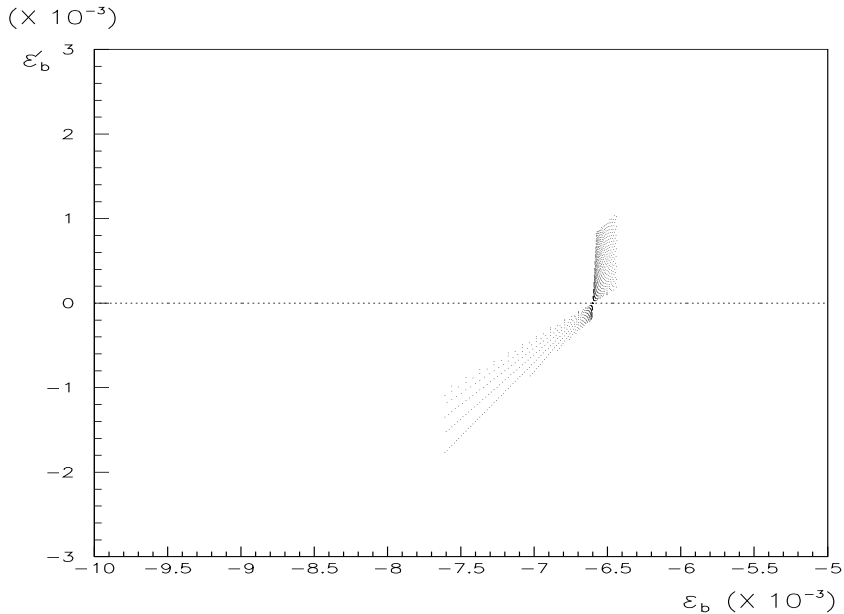,height=9cm,width=13cm}
\caption{ 
Allowed parameter sets of $(\epsilon_b,\epsilon'_b)$ in LR model
}
\vskip 0.5cm
\end{figure}

In this model we have two kinds of additional contributions
to the neutral current sector with right-handed current,
one is the interaction to extra neutral gauge boson $Z'$
and the other is the interaction to the ordinary $Z$ boson
suppressed by the mixing angle $\xi$. 
Since the $e^+ e^-$ collisions of LEP experiment
arise at the $Z$--peak energy,
the latter is relevant for our analysis,
which is given by
\bea
L_{NC} &=& \frac{e}{s_W c_W} Z^\mu \bar{b} \gamma_\mu
\left[ \left( T_3^L - s_W^2 Q + \xi t_R s_W ( T_3^L-Q) \right) P_L 
\right.
\nonumber \\
&&~~~~~~~~+ \left. \left( - s_W^2 Q 
      + \xi (s_W (t_R+\frac{1}{t_R}) T_3^R - t_R s_W Q) \right) P_R 
\right] b~,
\eea
where $t_R = \tan \theta_R$, $s_W = \sin \theta_W$, 
$c_W = \cos \theta_W$, and $Q$ is the electric charge.
$\xi$ is the neutral mixing angle between $Z$ and $Z'$.
The definitions of mixing angles follow the notation 
of Ref. \cite{chay}.
Since we do not impose a discrete L-R symmetry,
the additional parameter $\theta_R$ have to come into the model
besides $\xi$ and $m_{Z'}$.
The correction terms defined in Eq. (1) are expressed 
in terms of the model parameters:
\bea
\Delta^L_b &=& \Delta^{SM}_b (m_t^2) - \frac{1}{6} \xi t_R s_W
\nonumber \\
\Delta^R_b &=& - \frac{1}{6} \xi t_R s_W 
               \left( 1+\frac{3}{t_R^2} \right)~.
\eea
Note that the mass of extra $Z'$ boson does not enter the analysis 
for LEP I data since the LEP I experiment is performed
at the $Z$ peak energy.

\begin{figure}[t]
\epsfig{file=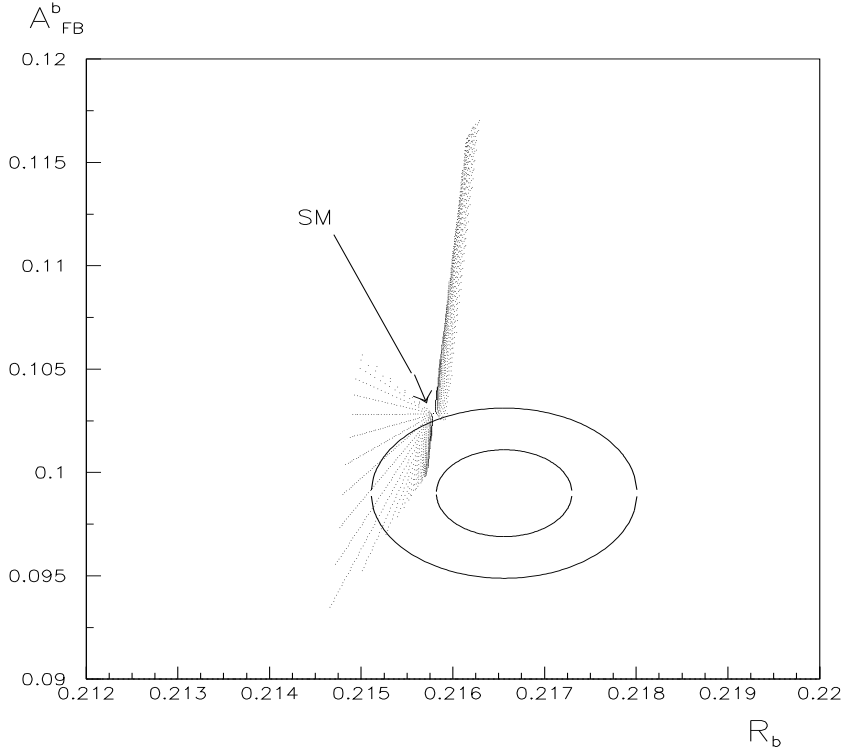,height=11cm,width=13cm}
\caption{ The LR model predictions of $R_b$ and $A_{FB}^b$ 
with the constrained values of $\epsilon$'s.
The inner ellipse denotes the experimental data at 1-$\sigma$ level 
and the outer ellipse at 95 $\%$ C.L..
}
\vskip 0.5cm
\end{figure}

When we consider the LR model,
the anomalous right-handed couplings to $Z$ boson
also appear in other fermion sectors.
Thus we have the heavy constraints on the mixing angle 
$\xi$ depending on $\theta_R$ from the lepton sectors 
\cite{chay,langacker,barenboim}.
In the analysis of the neutral sector of the general LR model 
in Ref. \cite{chay},
the constraints on $(\xi, \theta_R)$ are obtained 
from the $\epsilon_1$ and $\epsilon_3$ parameters using LEP data.
With these constraints on $\xi$ and Eq. (9), 
$\epsilon_b$ and $\epsilon'_b$ are also heavily constrained
as shown in Fig. 4.
Actually extremely small region out of the ellipse in the
$(\epsilon_b,\epsilon'_b)$ plain shown in Fig. 2
is consistent with the allowed $(\epsilon_1,\epsilon_3)$ values
in this model.
With these values of $\epsilon$'s,
the predictions of $A_{FB}^b$ and $R_b$ from the Eq. (9)
are presented in Fig. 5 together with the experimental data.
We find that there exist parameter sets with which the recent data
can be consistent with the LR model predictions at 95 $\%$ C.L..

\section{Concluding Remarks}

We consider the anomalous couplings of $Z b \bar{b}$ vertex
to explain the discrepancy in $A_{FB}^b$ 
as a manifestation of new physics effects.
Since there is no right--handed electroweak radiative corrections 
to the $Z b \bar{b}$ vertex in the large $m_t$ limit of the SM,
the anomalous right--handed currents of $Z b \bar{b}$ vertex
may be a sensitive probe to the new physics beyond the SM. 
Among the LEP observables, $A_{FB}^b$ is one of the best
window to explore the anomalous right--handed currents
since it is sensitive to the ratio of the right-handed coupling
to the left-handed coupling.

The electroweak precision variables are used 
to extract the new physics effects.
Any new physics effects in $Z l^+l^-$ vertex
are included in the $\Delta \rho$ and $\Delta k$
and consequently in $\epsilon_1$ and $\epsilon_3$.
Thus the corresponding relations of $\Gamma_l$, $A_{FB}^l$
are unchanged with the new physics effects.
Including the observable $A_{FB}^b$, we introduce a new variable 
$\epsilon'_b$ to probe the anomalous right-handed current
interactions in $Z b \bar{b}$ vertex.
Hence the new physics effects in $Z b \bar{b}$ vertex
are encoded in $\epsilon_b$ and $\epsilon'_b$ likewise.
From the experimental data we obtain the model independent bound
on $\epsilon_b$ and $\epsilon'_b$.
Note that the allowed parameter set 
$(\epsilon_1,\epsilon_3)$ should be altered
when we consider another set of observables.
However we find that the change is slight here 
since we introduce $A_{FB}^b$ and 
exclude $m_{_W}$ and their error pull is 
of the same order.

We consider the nonuniversal contact interactions and 
the LR model as underlying physics of 
such an anomalous right-handed current. 
The lower bound of the new physics scale
at which the higher dimensional operators arise
is estimated through the $\epsilon$ analysis.
The LR model provides shifts of $Z l^+ l^-$ couplings
as well as those of $Z b \bar{b}$ vertex
and they are closely related.
Thus the allowed region for $(\epsilon_b,\epsilon'_b)$ set
is constrained by the $(\epsilon_1,\epsilon_3)$ constraints
obtained from the precise measurements
of the leptonic current interactions.
The large amount of deviation in $A_{FB}^b$
prefers much shift of couplings
while the data of lepton sectors are rather closer 
to the SM predictions.
Therefore the combined analysis may present a strong constraint
on the neutral mixing angles of LR model
if the improved analyses of $A_{FB}^b$ would be performed.

\begin{ack}
This work is supported 
by the Korea Science and Engineering
Foundation through the SRC program of Center for 
Theoretical Physics at Seoul National University.
\end{ack}

\end{document}